\documentclass[reprint,amsmath,amssymb,groupedaddress]{revtex4-1}

\usepackage[english]{babel}

\usepackage{graphicx} 

\newcommand{\vct}[1]{\mathbf{#1}}
\newcommand{\kb}[0]{k_{\textup{B}}}

\usepackage{amsmath}
\usepackage{amssymb}
\usepackage{mathrsfs}

\begin{document}

\title{Self-assembly and cooperative dynamics of a model colloidal gel network}
\author{Jader Colombo}
\affiliation{Department of Civil, Environmental and Geomatic Engineering,
  ETH Z\"{u}rich, CH-8093 Z\"{u}rich, Switzerland}
\author{Emanuela Del Gado}
\affiliation{Department of Civil, Environmental and Geomatic Engineering,
  ETH Z\"{u}rich, CH-8093 Z\"{u}rich, Switzerland}

\begin{abstract}
We study the assembly into a gel network of colloidal particles, via effective
  interactions that yield local rigidity and make dilute network structures mechanically stable. The
  self-assembly process can be described by a Flory-Huggins theory, until a network of chains forms,
  whose mesh size is on the order of, or smaller than, the persistence length of the chains.  The
  localization of the particles in the network, akin to some extent to caging in dense glasses, is
  determined by the network topology, and the network restructuring, which takes place via bond
  breaking and recombination, is characterized by highly cooperative dynamics. We use $NVE$ and
  $NVT$ Molecular Dynamics as well as Langevin Dynamics and find a qualitatively similar time
  dependence of time correlations and of the dynamical susceptibility of the restructuring gel. This
  confirms that the cooperative dynamics emerge from the mesoscale organization of the network.
\end{abstract}

\maketitle

\section{Introduction}
Gels are disordered elastic materials that can form also via reversible aggregation and even in
extremely diluted particles suspensions~\cite{trappe_nature2001, plu_nature}. Their heterogeneous
microscopic structure---a particle network holding the solvent therein---is the result of the
complex interplay of aggregation, phase separation and arrested kinetics and the stress transmission
through it under deformation is far from trivial. Although being able to design the smart mechanics
of these fascinating materials at the level of their nanoscale components could be ground-breaking
for several technological
applications~\cite{di2013multistep,sacanna2013shaping,pochan_softmatter2010,barbara}, there is still
little theoretical understanding of the physical mechanisms underlying the development of their
mechanical response. The elastic moduli and the mechanical strength are not only determined by the
solvent-mediated effective interactions between the primary particles, but are also affected by the
mesoscale organization of the network and its topology: local rigidity, connectivity and degree of
heterogeneity of the structure can vary a lot depending on the system and on the arrest conditions
that eventually lead to gelation.  Various internal deformation modes, in particular the soft ones,
depending on the network topology, may dominate the elastic
response~\cite{alexander,Harrowell_NP08,wyart2008elasticity}. The nonlinear behavior is also a
hallmark of the mechanics of these materials~\cite{laurati_jor2011,everaers2000}. Nonaffine
microscopic deformations of the weakly connected structure certainly play an important role, but
here also the possible restructuring of the gel network has to be taken into account. It has been
suggested that breaking of weakly connected parts of the gel structure may be determinant for the
nonlinear response \cite{petekidis_softmatter2011, laurati_jor2011} and that their sudden breaking
or recombination produces its structural aging \cite{bouchaud,
  cipelletti2003_faraday}. Investigations of the microscopic dynamical processes indicate slow
cooperative dynamics very similar to the ones of dense glassy systems
\cite{cipelletti2003_faraday,duri2006length,af_jstat08,edgkob_prl07,kob_sastry_09} and the dynamical
heterogeneities observed in various studies have also been connected to several aspects of the
formation of the gel structure \cite{edgkob_jnnfm08,edgkob_epl05}. The link between structural
features, microscopic dynamical processes and mechanical response still needs to be established.

In recent years a number of approaches for simulating colloidal gelation have been proposed,
including short-range isotropic
interactions~\cite{testard2011,charbonneu2007_pre,tanaka2007_epl,foffi2005_jcp}, valence-limited and
patchy-particle models~\cite{zaccarelli2006gel,sciortino2011patchy,rovigatti2011}, dipolar
particles~\cite{blaak2007reversible}, and anisotropic effective
interactions~\cite{delgado2010microscopic,jaderema_prl,kob_sastry_09}. We follow here this latter
approach, using an effective interaction that includes, in the form of a three-body term, the basic
ingredients for a minimal model of particle gels. Using this approach, we were able to show in a
previous work that bond-breaking events in physical gels strongly couple to the complex network
topology and result in cooperative dynamics that are absent in chemical gels, in which bonds are
permanent~\cite{jaderema_prl}. This is due to the fact that bond breaking has nonlocal consequences,
because the network structure entails long range correlations: bond breaking is more likely to occur
in specific parts of the network, i.e.\ its soft parts, but induces an enhancement of particle
mobility and significant particle rearrangements relatively further away. We expect these features
of restructuring networks to play a crucial role in the mechanical response of physical gels.

In the present paper we thoroughly analyze the self-assembly process and the spatial structure of
the networks obtained within our model, and we present two main contributions concerning the
relaxation dynamics of colloidal gels. First, we characterize the cooperative dynamics of the gels
by means of the dynamical susceptibility $\chi_4$. This quantity has been extensively analyzed in
numerical simulations of dense glassy systems~\cite{dh}, but it has been relatively less studied in
gels~\cite{cipelletti2003_faraday,duri_pre-2005,duri2006length,abete_prl2007,coniglio2008dynamical,abete_pre2008},
in spite of the fact that experimental measures of $\chi_4$ in colloidal gels have already been
performed~\cite{duri_pre-2005,duri2006length}.  We compute $\chi_4$ and its dependence on the
scattering wave vector $q$ in networks with varying density, discussing possible connections with
experimental observations. Second, we assess to what extent the cooperative dynamics of the gel are
affected by the choice of the microscopic particle dynamics employed in the simulation (Newtonian
vs.\ stochastic, with different conserved quantities). In fact, in simulations of dense glasses it
has been shown that the amplitude of the dynamical fluctuations depends both on the statistical
ensemble and on the microscopic particle dynamics adopted~\cite{dh2}. We address this problem in our
model gel and confirm that the measured dynamical fluctuations, i.e.\ the peak amplitude of
$\chi_4$, depend on the microscopic particle dynamics, but the \emph{average} long-time dynamics do
not. Nonetheless, the scaling behavior of the peak with the scattering wave vector $q$ is robust
with respect to changes in the microscopic dynamics. Overall, we provide evidence that the
cooperative character of the long-time rearrangements in the gel is determined by its mesoscopic
network structure and is thus a feature appearing irrespective of the nature of the underlying
microscopic dynamics.

The paper is organized as follows. In Sec.~\ref{sec:numsim} we describe our model for colloidal gel
networks and provide the details of the numerical simulations we performed. In
Sec.~\ref{sec:structure} we analyze the self-assembly process through which the particles aggregate
into stress-bearing percolating networks, and characterize the structure of the resulting
gels. Section~\ref{sec:dynamics} is devoted to the analysis of the average particle dynamics and
relaxation processes. Section~\ref{sec:coopdyn} deals with dynamical fluctuations and cooperative
dynamics. In Sec.~\ref{sec:microdyn} we analyze the dependence of our findings on the microscopic
particle dynamics chosen for the numerical simulation. Finally, in Sec.~\ref{sec:conclusion} we
present a summary of the main results and some concluding remarks.

\section{Model and numerical simulations}\label{sec:numsim}

Dilute colloidal gels are characterized by open and thin network structures, where particle
coordination can be very low (two to three)~\cite{solomon,royall}. Hence the network connections need to
be fairly rigid to support at least their own weight and experiments have proven that bonds between
the colloidal particles can indeed support significant torques~\cite{pantina2005elasticity}. We
consider these as the two basic ingredients for a minimal model of particle gel networks, in the
same spirit as recent work~\cite{delgado2007length,saw2011computer,blaak2007reversible}, where
anisotropic interactions promote the assembly of thin open structures and stabilize them at low
volume fraction.

Our model system consists of $N$ identical particles with mass $m$, interacting via the potential
\begin{equation}\label{equ:poten}
U(\vct{r}_1,\ldots,\vct{r}_N) = \epsilon \left[\sum_{i > j} u_2\left(\frac{\vct{r}_{ij}}{\sigma}\right) +
 \sum_i\sum_{\substack{j>k}}^{j,k\ne i}u_3\left(\frac{\vct{r}_{ij}}{\sigma},\frac{\vct{r}_{ik}}{\sigma}\right)\right]\,,
\end{equation}
where $\vct{r}_{ij}=\vct{r}_j-\vct{r}_i$, $\vct{r}_i$ being the position vector of particle number
$i$, $\epsilon$ sets the energy scale, and $\sigma$ represents the particle diameter. Typical values
for a colloidal system are $\sigma = 10-100\,\rm{nm}$ and $\epsilon = 1-100\,\kb T_{\rm r}$, $\kb$
being the Boltzmann constant and $T_{\rm r}$ the room
temperature~\cite{trappe_nature2001,luca_faraday2003,laurati_jor2011}. The two-body term $u_2$
consists of a repulsive core complemented by a narrow attractive well:
\begin{equation}\label{equ:u2}
  u_2(\vct{r})=A\left(a\,r^{-18}-r^{-16}\right)\,.
\end{equation}
The three-body term $u_3$ confers angular rigidity to the interparticle bonds and prevents the
formation of compact clusters:
\begin{equation}
  u_3(\vct{r},\vct{r}') = B\,\Lambda(r) \Lambda(r')\,
  \exp\left[-\left(\frac{\vct{r}\cdot\vct{r}'}{rr'}-\cos\bar{\theta}\right)^2/w^2\right].
\end{equation}
The intensity of the the three-body interaction decays with increasing distance from the central
particle in accordance with the radial modulation
\begin{equation}
\Lambda(r)=
\begin{cases}
r^{-10} \left[1-(r/2)^{10}\right]^2 & r<2\,, \\
0 & r\ge 2\,.
\end{cases}
\end{equation} 

\begin{figure}
  \includegraphics[width=1.0\columnwidth]{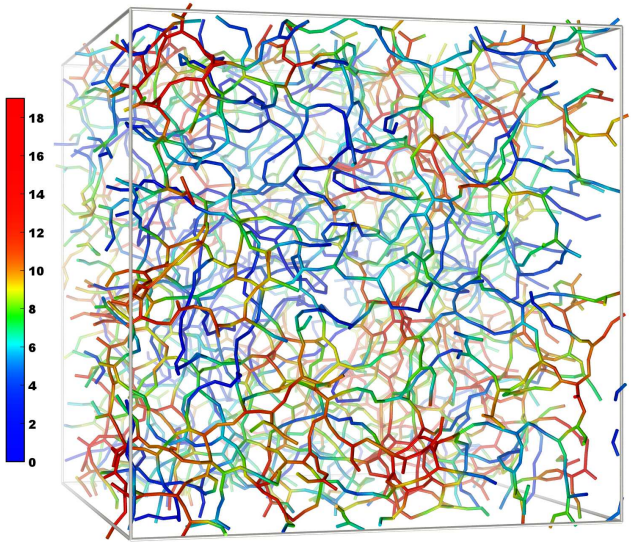}
  \caption{A snapshot of the particle network forming at number density $\rho=0.10$ (corresponding
    to an approximate volume fraction $\phi=5\%$) and temperature $T=0.05$. In order to visualize
    the network structure of the gel we do not show the particles, but instead represent the
    interparticle bonds with segments. The color code corresponds to the local density of crosslinks
    $c_i^{l_0}$ (see the text for the definition) highlighting the presence of strongly connected
    domains (red) and weakly connected domains (blue).\label{fig:partsnap}}
\end{figure}

The potential energy~\eqref{equ:poten} depends on the dimensionless parameters
$A,a,B,\bar{\theta},w$. In our study the model is not tailored to any specific colloidal system: we
have chosen these parameters such that for $\epsilon$ of the order of $10\kb T$ the particles start
to self-assemble into a persistent particle network like the one represented in
Fig.~\ref{fig:partsnap}. The data here discussed refer to $A=6.27, a=0.85, B=67.27,
\bar{\theta}=65^\circ, w=0.30$, one convenient choice to realize this condition.

In the following we will express distance in units of $\sigma$, energy in units of $\epsilon$, time
in units of $\sqrt{\sigma^2 m /\epsilon}$, and temperature in units of $\epsilon/\kb$. 

We performed molecular dynamics simulations with the LAMMPS source code~\cite{plimpton1995fast},
that we have suitably extended to include the potential~\eqref{equ:poten}. We used a cubic
simulation box with periodic boundary conditions. We have investigated systems with number density
$\rho=N/V=0.20$, $0.15$, $0.10$, and $0.05$; these correspond to approximate particle volume
fractions $\phi=10\%$, $7.5\%$, $5\%$, and $2.5\%$, respectively. The simulations we will report in
the following were performed with 16384 particles unless stated otherwise. We performed simulations
in the canonical ($NVT$) ensemble using the velocity Verlet algorithm coupled to a chain of
Nos\'e-Hoover thermostats, with a timestep $\Delta t=0.005$; a typical production run consisted in
$2\cdot10^8$ steps. In order to equilibrate the network we started at high temperature ($T=5.0$) and
cooled the system to a prescribed lower temperature over the course of $2\cdot10^7$ timesteps;
afterwards, the temperature was kept fixed at the target value. In the range of temperatures we will
report, structural and dynamical quantities did not show any aging with time after equilibration.
In order to investigate the role of different microscopic dynamics in our model gel we have also
performed simulations in the microcanonical ($NVE$) ensemble, and with Langevin dynamics, using in
both cases the same timestep as for $NVT$ dynamics. For the simulations in the $NVE$ ensemble we
first equilibrated the system at the target temperature with the Nos\'e-Hoover thermostat; we then
evolved the system at constant energy for $2\cdot10^8$ steps. In the case of Langevin dynamics we
used a drag force $\vct{F}_i=-m\gamma \dot{\vct{x}}_i$ with a drag coefficient $\gamma=10.0$, and a
typical production run covered $6\cdot10^8$ steps.

\begin{figure}
  \centering
  \includegraphics[width=1.\columnwidth, clip]{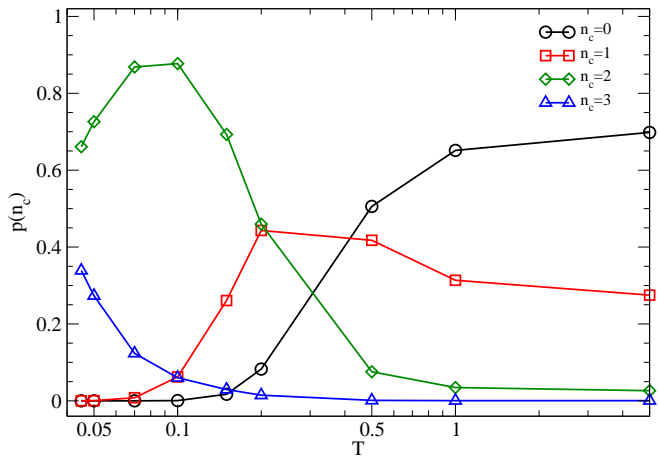}
  \caption{Fraction of particles having $n_c$ bonds (ranging from zero to three) as function of
    temperature, in a gel network with volume fraction $\phi=7.5\%$. \label{fig:connectivity}}
\end{figure}

In the range of temperatures of interest here, interparticle bonds are not permanent: thermal
fluctuations favor breaking of existing bonds and formation of new ones between particles
that were previously nonbonded. In order to identify how these processes affect the network dynamics 
we have also performed simulations where we added a narrow potential barrier
$u_2^c(r)$ to the two-body term~\eqref{equ:u2} of the interaction potential:
\begin{equation}
  u_2^c(r) = C\exp\left[-(r-r_0)^2/\delta^2\right]\,,
\end{equation}
with $C=10.0$, $r_0=1.2$, and $\delta=0.01$. This enabled us to switch at will, for the same gel,
between a \emph{restructuring} network, in which bonds may be broken and created, to a
\emph{nonrestructuring} one, where the particle connections do not change over
time~\cite{zaccarelli_barrier,kob_sastry_09}.

\section{Self-assembly, spatial correlations and network structure}\label{sec:structure}



Upon lowering the temperature the particles progressively assemble into a network. Pairs of
particles separated by a distance $r \leq r_{\rm min}\approx 1.2 \sigma$, around which is centered
the first minimum of the radial distribution function, are considered bonded.\footnote[2]{Since we
  will be interested in detecting the breaking and the formation of bonds over the course of time,
  we find it convenient to adopt a bond criterion with hysteresis, which helps in reducing false
  positives due to vibrations of the bonds: two particles are considered bonded as soon as their
  distance becomes smaller than a first threshold $r_a=1.1$, but an existing bond is considered
  broken only when their distance becomes larger than a second threshold $r_b=1.3$. We stress that
  this is just a criterion used for the purpose of data analysis: there is no such distinction
  between bonded and nonbonded particles in the potential~\eqref{equ:poten}. All results discussed
  here are not sensitive to the specific choice of $r_a$ and $r_b$, as long as they bracket $r_{\rm
    min}$.} The average fraction of particles with $n_c$ bonds, $p(n_c)$, is shown as a function of
temperature in Fig.~\ref{fig:connectivity}.  At high temperature most particles do not form any
long-lived bond and exist as monomers ($n_c=0$), although there is a certain fraction of transient
dimers ($n_c=1$). Upon lowering the temperature chains with an increasing number of particles begin
to form, so that the fraction of two-coordinated particles increases significantly. We can
distinguish a first regime where $p(n_c=2)$ increases while $p(n_c=1)$ is roughly constant. Below
$T\simeq 0.2$ chains grow at the expense of dimers and in the region $T\lesssim 0.1$ the chains
branch, with a notable growth of three-coordinated particles (crosslinks or {\it nodes}) at the
expense of two-coordinated particles.
\begin{figure}
  \centering
      \includegraphics[width=1.\columnwidth, clip]{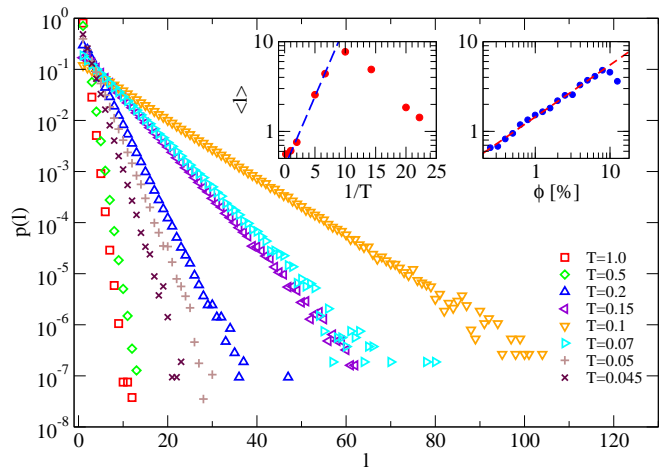}
  \caption{(Main plot) Chain length distribution $p(l)$ for a set of temperatures in a gel network
    with particle volume fraction $\phi=7.5\%$. (Left inset) Average chain length as a function of
    inverse temperature at fixed volume fraction $\phi=7.5\%$; the dashed line represents the
    relation $\langle l \rangle\sim \exp[E/(2T)]$, with $E\approx 0.7$. (Right inset) Average chain
    length as a function of volume fraction at fixed temperature $T=0.15$; the dashed line
    represents the relation $\langle l \rangle\sim\phi^{0.6}$. \label{fig:chains}}
\end{figure}

Having defined the chains as clusters of bonded one- or two-coordinated particles, the fraction of
chains with length $l$, $p(l)$, is plotted in Fig.~\ref{fig:chains} (main plot) for different
temperatures.  For $T \geq 0.1$, i.e.\ before the onset of branching, we expect the chain length
distribution to be well described by a Flory-Huggins (mean field) type of theory
\cite{degennes1979scaling,cates1990statics}, and hence that at relatively low volume fractions the
chain length follows an exponential law $p(l)\sim\exp[-l/\langle l \rangle]$, with an average
$\langle l \rangle$ that increases with lowering the temperature or increasing the density as
$\langle l \rangle \sim \exp[E/(2T)]\phi^y$, in which $E$ is the scission energy of a chain and
$y\approx 0.6$. Our data is compatible with the expected dependence on both the temperature (left
inset of Fig.~\ref{fig:chains}) and density (right inset).  When branching of chains comes into
play, i.e.\ around $T\simeq 0.1$ in the left inset and $\phi\simeq 7.5\%$ in the right inset, the
average chain length becomes instead progressively smaller with decreasing temperature or increasing
density, respectively.

\begin{figure}
  \includegraphics[width=1.0\columnwidth, clip]{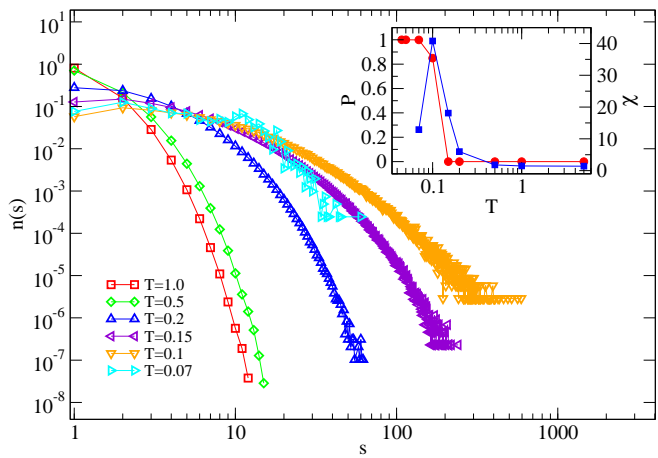}
  \caption{(Main plot) Cluster size distribution $n(s)$ for $\phi=7.5\%$ and a set of
    temperatures. The distribution is normalized so that $\sum_{s=1}^\infty n_s = 1$. (Inset) Blue
    squares: mean cluster size $\chi$ as a function of temperature. Red circles: $P$, fraction of
    particles belonging to a percolating cluster, as a function of temperature.\label{fig:cs}}
\end{figure}

To investigate also the regime where branching or crosslinking of the chains becomes important, we
define more generally clusters that are maximal connected subsets of particles. The fraction $n(s)$
of clusters made of $s$ bonded particles significantly deviates, at low temperatures, from the
exponential size distribution typical of transient, short-lived clusters formed upon particle
collisions. It is shown in Fig.~\ref{fig:cs} for $\phi=7.5\%$ and different temperatures.
We also consider that a cluster is percolating if contains a path connecting a particle to one of
its periodic images. From the high-temperature exponential law indicating the formation of
short-lived clusters upon particle collisions, by lowering the temperature the cluster size
distribution becomes nonmonotonic, with dimers having a higher statistical frequency than monomers,
and eventually develops a power-law tail, compatible with a percolation
phenomenon~\cite{stauffer1994book}.  The fraction $P$ of particles belonging to a percolating
cluster shows a sudden increase when the temperature drops below $0.1$, where the mean cluster size
$\chi$, defined as $\sum_{s=1}^{\infty}s^2 n_s/\sum_{s=1}^{\infty}s n_s$, also displays a peak
(inset of Fig.~\ref{fig:cs}). For $T<0.1$ practically all particles (more than $99\%$) belong to a
single percolating cluster, that is a persistent, connected network of particles spanning the whole
system.

\begin{figure}
  \begin{center}
      \includegraphics[width=1.\columnwidth, clip]{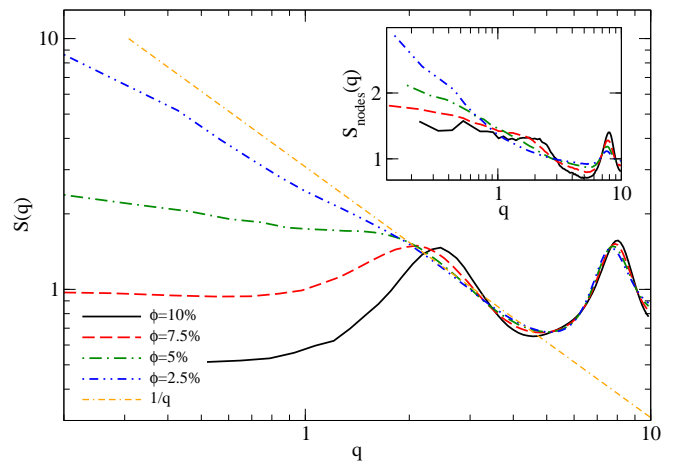}
  \end{center}
  \caption{Structure factor $S(q)$ of the gel (main plot) and of the nodes alone (inset) for
    different volume fractions at fixed temperature $T=0.05$.
    \label{fig:strfact_rho}}
\end{figure}

In order to characterize the structure of the network over different length scales we
compute the static structure factor
\begin{equation}
  S(q) = \left\langle\frac{1}{N}\sum_{j,k=1}^N \exp[i\vct{q}\cdot(\vct{r}_j-\vct{r}_k)]\right\rangle
\end{equation}
that quantifies spatial correlations between particle positions over distances $\simeq 2 \pi /q$
(where $q$ is in units of $\sigma^{-1}$), regardless of whether they are connected or not through
the network.  The high-temperature $S(q)$ is essentially featureless, but upon lowering the
temperature it develops patterns that allow to follow the network development. In
Fig.~\ref{fig:strfact_rho} we plot $S(q)$ computed in networks obtained at $T=0.05$ for different
volume fractions. The peak at $q_b\simeq 8$ corresponds to distances of the order of the typical
bond length and hence quantifies the contribution of bonded particles. This is weakly dependent on
density within the range investigated. The peak at $q^*\simeq 2$ signals spatial correlations
between particles separated by distances $\simeq 3$-$5\sigma$.  These and, even more, the
correlations over larger length scales are instead significantly affected by density.  For
intermediate wave vectors the $q^{-1}$ behavior expected for linear chains is approached. The
departure from this regime shifts to lower wave vectors upon lowering the volume fraction, signaling
an increase in the average chain length.

\begin{figure}
  \begin{center}
      \includegraphics[width=1.\columnwidth, clip]{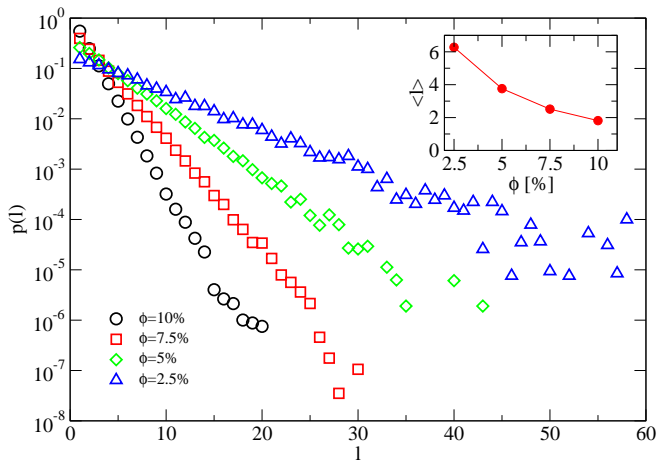}
  \end{center}
  \caption{(Main plot) Chain length distribution $p(l)$ for different volume fractions at fixed
    temperature $T=0.05$. (Inset) Average chain length $\langle l \rangle$ as a function of
    volume fraction. \label{fig:chains_rho}}
\end{figure}

To clarify the origin of the {\it mesoscale} correlations and in particular of the peak at $q^*$ we
compute the mesh size of the network by evaluating the distribution $p(l)$ of chain lengths between
nodes, which is plotted in Fig. \ref{fig:chains_rho} for various volume fractions. From the
exponential distributions we extract the average chain length $\langle l\rangle$ that gives us an
estimate of the average mesh size of the network, shown in the inset. The data show that both the
value of $\langle l \rangle$ and its dependence on $\phi$ are consistent with the position of the
peak at low $q$ in the structure factor.

\begin{figure}
  \begin{center}
      \includegraphics[clip=true,width=1.\columnwidth]{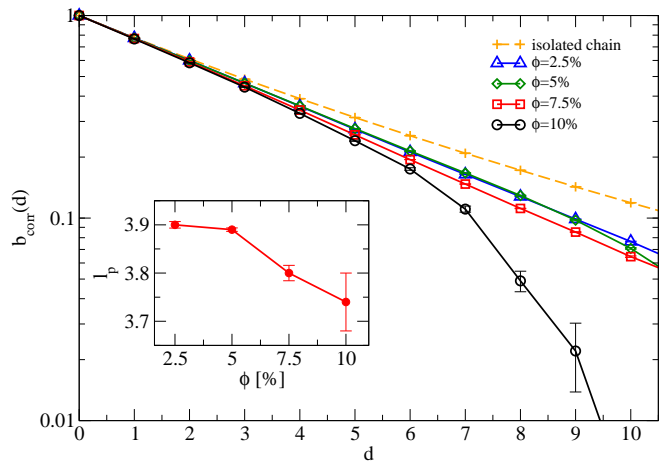}
  \end{center}
  \caption{(Main plot) Bond orientational correlation function $b_{\rm corr}(d)$ for an isolated
    chain and for chains embedded in networks with different volume fractions. (Inset) Persistence
    length, obtained by fitting an exponential function to $b_{\rm corr}(d)$ in the interval $0\le
    d\le5$, as a function of volume fraction.\label{fig:perlen}}
\end{figure}

 In addition to the network connectivity, another origin of mesoscale correlations in the network
 could be the local rigidity of the chains. We estimate the persistence length of the network chains
 by introducing an orientational correlation function of the chain bonds, as explained in the
 following. Considering a chain of length $l$ in a given network configuration, we denote by
 $[\pi_0, \pi_1,\ldots,\pi_l]$ the sequence of particle indices corresponding to the sequence of
 particles forming the chain. The $i$-th particle-particle bond in the chain is then $\vct{b}_i =
 \vct{r}_{\pi_{i+1}}-\vct{r}_{\pi_i}$, with $0\le i <l$. The correlation function $b_{\rm corr}(d)$
 measures the coherence between the orientation of two bonds separated by $d$ particles along the
 same chain; it is defined by $b_{\rm corr}(d)=\langle\vct{b}_0\cdot\vct{b}_d\rangle$, the angular
 brackets denoting an average over (i) the set of all chains in a given network configuration, and
 (ii) a sequence of configurations sampling the time evolution of the network. The range of the
 correlation function is restricted to the maximum chain length and is therefore density-dependent.
 The results are plotted in Fig.\ref{fig:perlen} and compared to the same bond orientational
 correlations obtained for long, isolated chains, i.e.\ chains not embedded in a network: from the
 exponential decay we obtain a persistence length $l_p\approx 4$ bonds (see inset).  Being
 $l_{p}\geq 2\pi/q^*$, we can conclude that correlations over these length scales must be also due
 to the persistence length of the chains, that therefore certainly contributes to the peak at $q^*$
 in the structure factor. Nevertheless, $l_p$ does not change significantly within the range of
 densities investigated here, therefore the fact that the position of the peak moves towards lower
 $q$ upon decreasing the density is instead controlled by the mesh size (or the length of the chains
 at higher temperatures, when the network has not formed yet). It is interesting to note that the
 persistence length of the chains in the networks is larger than, or of the order of, their contour
 length, indicating that this gel network could have features quite similar to semiflexible networks
 typical of biopolymer systems~\cite{fred_prl1, fred_prl2, cyron2013equilibrium}.

The contribution to $S(q)$ of the network nodes alone is shown in the inset of
Fig.~\ref{fig:strfact_rho}: the data indicates that a certain fraction of them are nearest or
next-nearest neighbors, creating densely connected regions. This is confirmed when we associate to
each particle $i$ a local density of nodes $c_i^{l_0}$ that measures the number of nodes lying
within a distance of $l_0=5$ bonds along the network: the spatial distribution of $c_i^{l_0}$ is
highly inhomogeneous, ranging from 0 in loosely connected regions to $\approx 20$ in strongly
connected ones~\cite{jaderema_prl}. The reader may appreciate the existence of regions
  with different density of crosslinks also by looking at the gel snapshot in
  Fig.~\ref{fig:partsnap}. The strong signal in the node $S(q)$ at small $q$ shows another common
feature of dilute networks, emerging from long-range spatial correlations between nodes across
loosely connected domains \cite{delgado2010microscopic}.

\section{Particle localization in the gel network}\label{sec:dynamics}

\begin{figure}
  \begin{center}
      \includegraphics[width=1.\columnwidth, clip=true]{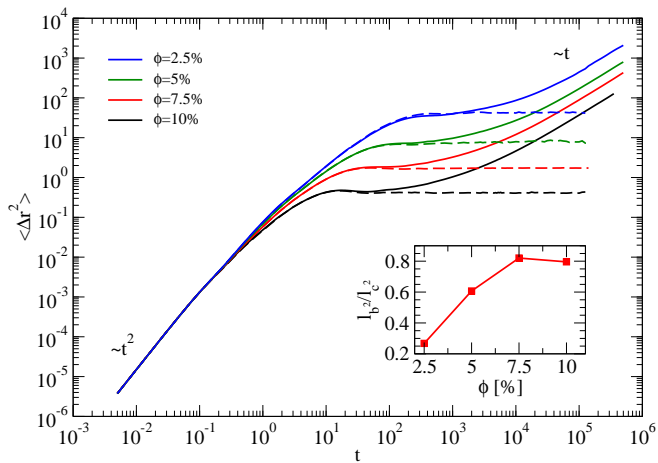}
  \end{center}
  \caption{(Main plot) Mean square displacement for different volume fractions at fixed temperature
    $T=0.05$ in the restructuring networks (full lines) and in the nonrestructuring networks (dashed
    lines). The volume fraction $\phi$ is, from top to bottom: 2.5\%, 5\%, 7.5\%, and 10\%. (Inset)
    Ratio between the average blob size and the average cage size as a function of volume
    fraction. \label{fig:msd}}
\end{figure}

We first characterize the average single-particle dynamics using the mean square displacement (MSD)
\begin{equation}
  \langle\Delta r^2(t)\rangle = \left\langle\frac{1}{N}\sum_{i=1}^N
  (\vct{r}_i(t)-\vct{r}_i(0))^2\right\rangle\,,
  \label{equ:msd}
\end{equation}
which we plot in Fig.~\ref{fig:msd} for $T=0.05$ and different volume fractions. In the
restructuring networks (full lines) $\langle\Delta r^2\rangle$ has a transition from a ballistic
behavior ($\sim t^2$) at short times (due to the Newtonian dynamics we use here) to a diffusive
behavior ($\sim t$) at long times. A pronounced plateau---corresponding to a time window in which
particles are transiently localized---separates the two regimes. The point of inflection of the mean
square displacement defines a density-dependent localization time $t_{\rm loc}$, which increases
from $t_{\rm loc} \approx 20$ at $\phi=10\%$ to $t_{\rm loc} \approx 400$ at $\phi=2.5\%$. The
corresponding localization length, defined as $l_c=\langle\Delta r^2(t_{\rm loc})\rangle^{1/2}$,
increases from $l_c\approx 0.6$ to $l_c\approx 6$. This type of particle localization is strongly
reminiscent of the {\it caging} in the dynamics of dense glasses~\cite{glasses}, but whereas there
this is due to the particles being embedded in a crowded environment, in the gel it is a consequence
of the particles being constrained by the network structure. Consequently, the localization length
of the particles can be much larger than the one typical of dense
glasses~\cite{zaccarelli2005model,cristiano_softmatter2011}.  Figure~\ref{fig:msd} in fact shows the
mean square displacements in the nonrestructuring networks (dashed lines), which up to $t_{\rm loc}$
coincide with the ones of the restructuring networks; after the localization time they are instead
constant and roughly equal to $l_c$. This shows that the localization length is mainly controlled by
the network topology (having fixed the effective interactions that stabilize the structure),
i.e.\ the average fraction of two- and three-coordinated particles. On the contrary, the long-time
diffusion of particles in the gel hinges on network restructuring.
\begin{figure}
  \begin{center}
      \includegraphics[width=1.\columnwidth, clip=true]{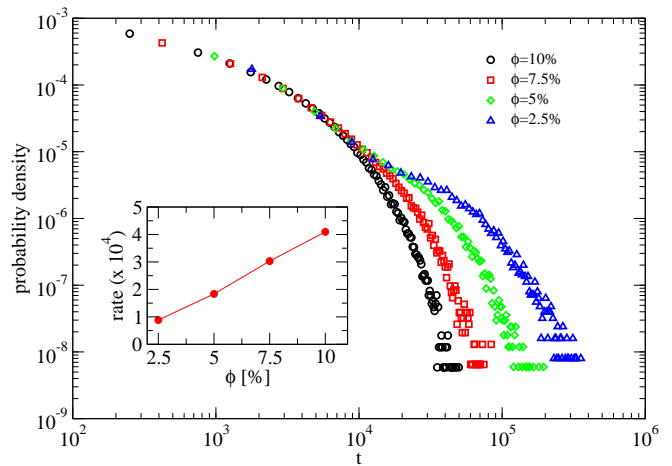}
  \end{center}
  \caption{(Main plot) Probability distribution of the time intervening between the formation of a
    bond and its subsequent breaking in networks with different volume fractions; the temperature is
    fixed at $T=0.05$. (Inset) Breaking rate (inverse of the average breaking time) as a function of
    the volume fraction. \label{fig:bbstat}}
\end{figure}
One might think that the escaping of a particle from its cage is a completely local process,
corresponding to the breaking of one or more bonds linking the particle to the rest of the network.
In order to test this hypothesis we partition the trajectory of each particle $i$, starting at time
$t_0=0$ and ending at time $t_f$ corresponding to the entire simulation window, into a set of
\emph{blobs}, on the basis of changes in the coordination number of the particle. 
  Whenever the particle acquires a new neighbor, or it looses an existing neighbor, the current blob
  ends and the next one starts: ${B}_1^i=[t_0^i=0,t_1^i],
  {B}_2^i=[t_1^i,t_2^i],\ldots,{B}_n^i=[t_{n-1}^i,t_n^i=t_f]$.  We define the size of a blob as the
  variance of the particle position over the corresponding time interval:
\begin{equation}
  \|{B}_k^i\|^2 = \langle\vct{r}^2_i(t)\rangle_{[t_{k-1}^i,t_k^i]} -
  \langle\vct{r}_i(t)\rangle_{[t_{k-1}^i,t_k^i]}^2\,,
\end{equation}
and compute the mean blob size $l_b^2$ by averaging over the blobs of all particles:
\begin{equation}
  l_b^2 = \left\langle\mathscr\|{B}_k^i\|^2\right\rangle_{i,k}\,.
\end{equation}
This mean blob size quantifies the extent of particle localization on the time scale typical of
changes in the local particle environment. In the inset of Fig.~\ref{fig:msd} we plot $l_b^2/l_c^2$
as a function of the volume fraction. The data show that the localization length measured in this
way in between bond-breaking events is of the order of the one detected by the MSD at volume
fractions $\simeq 10\%$, but becomes significantly smaller than that upon decreasing the volume
fraction. This suggests that the cage effect detected by the MSD should not be thought of in terms
of the two-three bonds linking the particle to the network, but rather in terms of a relatively
larger portion of the structure constraining the particle motion.

Useful information on the bond-breaking processes in the network can be obtained by computing the
probability distribution (i.e.\ normalized histogram) of the time lag between the formation of a
bond and its subsequent breaking, which is shown in Fig.~\ref{fig:bbstat}; the data refer to
$T=0.05$ and various volume fractions $\phi$. One can see that the long-time behavior of the
distribution varies significantly with $\phi$ , suggesting that bond breaking is not simply a
thermally activated process, but is instead significantly affected by the large scale structure of
the network, which changes with $\phi$, as indicated by the structure factor
(Fig.~\ref{fig:strfact_rho}).

\begin{figure}
  \begin{center}
      \includegraphics[width=1.\columnwidth, clip=true]{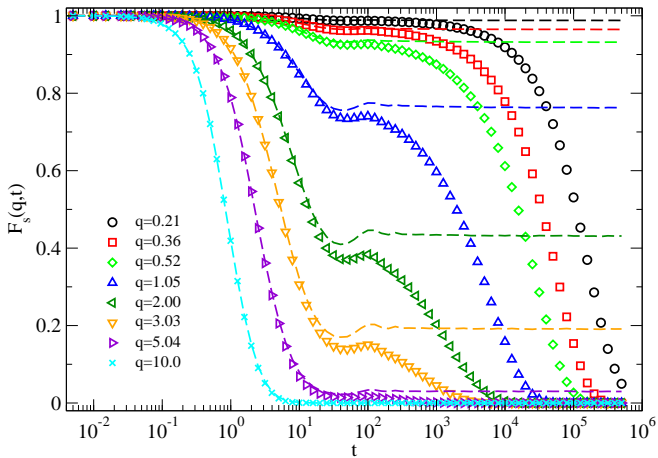}
  \end{center}
  \caption{Incoherent scattering function $F_s(q,t)$ in a restructuring network with volume
    fraction $\phi=7.5\%$ and temperature $T=0.05$ (symbols), and in the corresponding
    non-restructuring network (dashed lines). \label{fig:sisf}}
\end{figure}

We analyze now the spatial dependence of the single-particle dynamics by computing the incoherent
scattering function
\begin{equation}\label{equ:sisf}
  F_s(q,t) = \langle\Phi_s(q,t)\rangle \,,
\end{equation}
where $\langle\ldots\rangle$ indicates a time average and
\begin{equation}
  \Phi_s(q,t) = \frac{1}{N} \sum_{j=1}^{N}
  \exp\left[-i\vct{q}\cdot\left(\vct{r}_j(t)-\vct{r}_j(0)\right)\right]\,.
\end{equation}
This function quantifies the time correlation in single-particle displacements over a length scale
$\approx 2\pi/q$ and a time lag $t$. In Fig.~\ref{fig:sisf} we plot $F_s(q,t)$ for a set of wave
vectors at volume fraction $\phi=7.5\%$ and temperature $T=0.05$ in a restructuring network
(symbols) and in a nonrestructuring network (dashed lines) with the same topology. In the
restructuring network, for $q\lesssim 5$ the two-step decay of time correlations indicates a
separation of time scales (i.e.\ between a fast process and a slow process) with a plateau
corresponding to a localization regime around a time $t_{\rm loc}\approx 10^2$. This is again
strongly reminiscent of the caging phenomenon in the dynamics of dense glasses~\cite{glasses},
consistent with the time dependence of the MSD (Fig.~\ref{fig:msd}). Here we can see that the height
of the plateau decreases with increasing wave vector; at sufficiently high wave vector ($q\gtrsim5$)
the plateau disappears completely, indicating that the fast relaxation process is the only one
relevant at small length scales.

\begin{figure}
  \begin{center}
      \includegraphics[width=1.\columnwidth, clip=true]{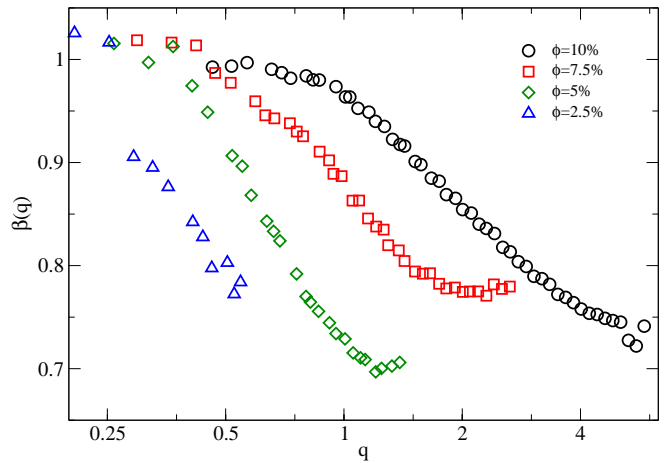}
  \end{center}
  \caption{Stretching exponent $\beta$ as a function of the wave vector $q$ in four networks with
    different volume fractions; the temperature is fixed at $T=0.05$. \label{fig:strexp}}
\end{figure}

The decay of correlations in the restructuring and nonrestructuring networks coincide up to $t_{\rm
  loc}$: therefore, the fast relaxation process is due to particle motion---network
``vibrations''---that does not entail breaking of bonds; this is somewhat similar to cage rattling in
glasses.  The decay of correlations after the plateau is instead due to network restructuring, since
it takes place only if bonds are allowed to break. This decay can be fitted by a stretched
exponential function
\begin{equation}\label{equ:strexp}
F_s(q,t) \sim \exp[-(t/\tau_q)^\beta]\,,
\end{equation}
where $\tau_q$ is the structural relaxation time and $\beta$ a $q$-dependent stretching
exponent. Our data indicate a strong dependence of the relaxation dynamics on the length scale set
by $q$. In Fig.~\ref{fig:strexp} we plot the stretching exponent as a function of the wave vector in
four networks with different volume fraction. At very large length scales (i.e.\ low wave vector)
the incoherent scattering function decays exponentially ($\beta \approx 1$) as one would expect for
simple particle diffusion. However, with increasing $q$ within the range where the plateau is
observed the decay of correlations becomes more and more stretched ($\beta<1$): in all cases $\beta$
decreases from unity down to $0.70\div0.75$. These results indicate that the decay of time
correlations changes qualitatively with the size of the region under analysis: in particular, there
is an intermediate range of wave vectors where relaxation dynamics become complex and slow. The
range of wave vectors where the stretched exponential decay is observed increases with increasing
volume fraction, suggesting that in networks obtained at higher $\phi$ the correlations in the
dynamics may be more extended. Following the model developed in Ref.~\onlinecite{krallweitz1998}, we
have used these data to extract the dependence of the localization length on the volume fraction:
within the relatively limited range of volume fractions investigated here, the results are
consistent with the prediction of Krall and Weitz. Our findings are also reminiscent of the results
of dynamic light scattering experiments conducted on colloidal gels and other jammed soft materials,
in which they report a two-step decay of dynamical correlators, a length scale dependence of the
slow relaxation dynamics, and a $\beta$ exponent decreasing with increasing wave
vector~\cite{cipelletti2000_prl, duri2006length, cipelletti2003_faraday}. However, while we find
that the decay of time correlations is exponential at low wave vector and becomes progressively
\emph{stretched} with increasing $q$, in the experiments a \emph{compressed} decay
($\beta\lesssim1.5$) is consistently reported. This behavior has been connected to microcollapses of
the gel structure under the action of internal stresses driving the aging of the
material~\cite{cipelletti2003_faraday, bouchaud}. The absence of aging in our model (at least on the
time scale accessible in the simulations) might explain why we do not observe any compressed
exponential behavior.

\begin{figure}
  \begin{center}
      \includegraphics[width=1.\columnwidth, clip=true]{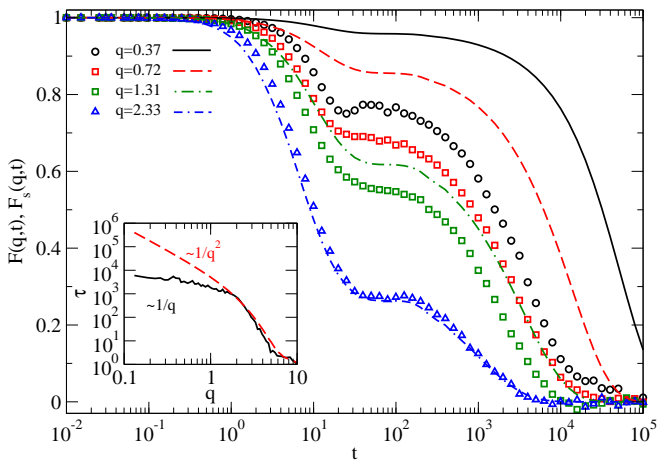}
  \end{center}
  \caption{(Main plot) Comparison between the coherent scattering function $F(q,t)$
    (symbols) and the incoherent part $F_s(q,t)$ (lines) in a restructuring network with volume fraction
    $\phi=7.5\%$ and temperature $T=0.05$, for a selection of wave vectors. (Inset) Structural
    relaxation time $\tau_q$ as determined from $F(q,t)$ (full line) and $F_s(q,t)$ (dashed
    line).\label{fig:cisf}}
\end{figure}

We have also computed the \emph{coherent} scattering function
\begin{equation}
  F(q,t)= \frac{1}{NS(q)}\sum_{j,k=1}^N \exp\left[-i\vct{q}\cdot\left(\vct{r}_k(t)-\vct{r}_j(0)\right)\right]\,,
\end{equation}
which provides information on the collective dynamics of the system and is akin to the intensity
autocorrelation function accessible in light scattering experiments. Figure~\ref{fig:cisf} shows a
comparison between $F(q,t)$ (symbols) and $F_s(q,t)$ (lines) for selected wave vectors in a gel
network with volume fraction $\phi=7.5\%$ and temperature $T=0.05$. The coherent scattering function
follows the same two-step decay pattern shown by the incoherent part and already discussed. At high
wave vector the two curves are on top of each other, whereas at low wave vector $F(q,t)$ decays
faster than $F_s(q,t)$. We have evaluated the $q$-dependent structural relaxation time of the
network by computing the area under the relaxation function: $\tau_q=\int_0^\infty
\mathscr{F}(q,t)dt$, where $\mathscr{F}(q,t)$ stands either for $F(q,t)$ or $F_s(q,t)$. The results
are shown in the inset of Fig.~\ref{fig:cisf}. At low wave vector the relaxation time deduced from
the incoherent scattering function scales as $q^{-2}$, suggesting that the large-scale
single-particle dynamics are diffusive. On the contrary, the relaxation time extracted from the
coherent scattering function has a ballistic scaling ($\tau_q\sim q^{-1}$), in accordance with
experimental results on gels and other soft materials~\cite{duri2006length, cipelletti2003_faraday}.

Overall our study of the average single-particle dynamics hints in several places to the fact that
such dynamics are strongly affected by the mesoscale organization of the network, and therefore
characterized by long-range correlations and presumably strong cooperativity. In the following we
quantify these correlations and analyze their origin.

\section{Cooperative dynamics}\label{sec:coopdyn}

\begin{figure}
  \begin{center}
    \includegraphics[width=1.\columnwidth, clip=true]{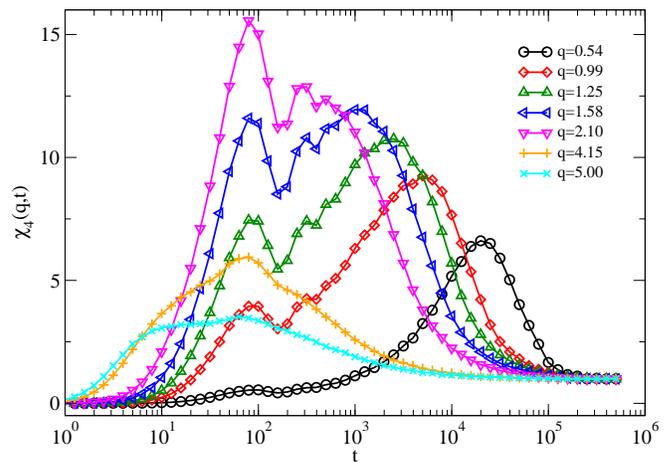}
  \end{center}
  \caption{Dynamical susceptibility $\chi_4(q,t)$ in a restructuring network with $\phi=7.5\%$,
    $T=0.05$. \label{fig:chi4}}
\end{figure}
\begin{figure}
  \begin{center}
    \includegraphics[width=1.\columnwidth, clip=true]{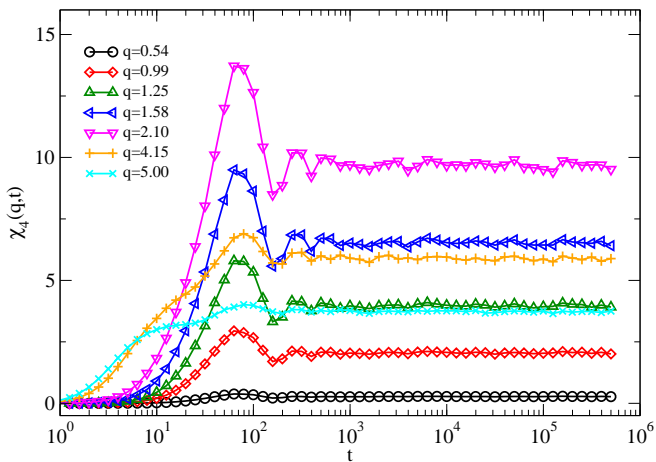}
  \end{center}
  \caption{Dynamical susceptibility for the same network as in Fig.~\ref{fig:chi4}, but in which the
    bonds have been constrained to prevent network restructuring. 
    \label{fig:chi4_cnstr}}
\end{figure}

We analyze the degree of cooperativity in the dynamics of the gel network by computing the variance 
\begin{equation}\label{equ:chi4}
  \chi_4(q,t) = N\left[\langle|\Phi_s(q,t)|^2\rangle-\langle\Phi_s(q,t)\rangle^2\right]\,.
\end{equation}
This dynamical susceptibility detects fluctuations from the mean degree of correlation in single
particle displacements due to spatial correlations of the dynamics, i.e.\ to particles undergoing
cooperative motion over a distance $\approx 2\pi/q$ and a time $t$~\cite{dh,dh1,dh3}. In
Fig.~\ref{fig:chi4} we plot $\chi_4$ as a function of time for a set of wave vectors in a
restructuring network with volume fraction $\phi=7.5\%$ and temperature $T=0.05$. At large
wave vectors the curves start from zero, have a peak corresponding to the localization time $t_{\rm
  loc}\approx 10^2$, then decay to the asymptotic value $\chi_4(q, t\to\infty) = 1$. The shape of
the curves changes qualitatively for $q\lesssim q^*$, $q^*\approx 2$ being the peak in the structure
factor corresponding to the mesh size of the network (see Section~\ref{sec:structure}).  In this
regime the curves display a second peak in correspondence of the structural relaxation time
$\tau_q$. Accordingly, the peak shifts at larger times with decreasing $q$; at the same time, the
amplitude of the peak diminishes. This is again very similar to what is found in dense glassy
suspensions~\cite{dh}.
Figure~\ref{fig:chi4_cnstr} shows the dynamical susceptibility for the same network as in
Fig.~\ref{fig:chi4}, but in which the bonds have been constrained so that no restructuring
happens. Since the dynamics of the two systems coincide until the localization time is reached,
$\chi_4$ displays the same features up to the first peak. Beyond $t_{\rm loc}$, however, the curves
are markedly different: the second peak is totally lacking in the constrained network; $\chi_4$
reaches instead a plateau value close to the amplitude of the first peak and therefore different
from the asymptotic value attained in the restructuring case (unity), as it is expected for chemical
gels \cite{abete_prl2007,abete_pre2008,goldbart1}.
The comparison between the two networks 
indicates that whereas the first peak has to be ascribed to vibrational motion present in both systems, 
the second peak is associated solely to the restructuring process. 
\begin{figure}
 \begin{center}
   \includegraphics[width=1.\columnwidth, clip=true]{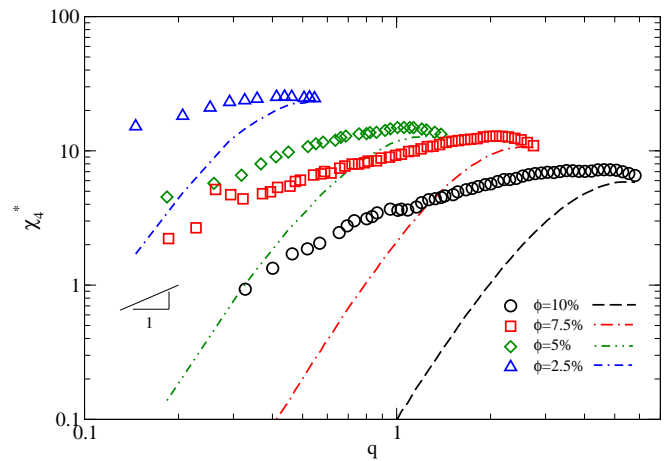}
 \end{center}
  \caption{Amplitude of the restructuring peak of the dynamical susceptibility, $\chi_4^*$, as a
    function of the wave vector $q$ in four networks with different volume fractions (symbols),
    compared to the plateau value of $\chi_4(q,t)$ in the corresponding nonrestructuring networks
    (lines). The temperature is $T=0.05$. \label{fig:chi4star}}
\end{figure}

We concentrate now on the peak in the dynamical susceptibility associated to network restructuring,
and whose $q$-dependent amplitude is $\chi_4^*(q)$. In Fig.~\ref{fig:chi4star} we plot $\chi_4^*$ as
a function of the wave vector in networks assembled at different volume fractions (symbols). The set
of wave vectors shown for each density corresponds to the $q$-range where in the restructuring case
we are able to identify the second peak in $\chi_4(q,t)$, which also coincides with the range where
$F_s(q,t)$ displays a stretched exponential decay (see Fig.~\ref{fig:strexp}). In the same plot the
plateau value reached by $\chi_4(q,t)$ for the same but nonrestructuring networks is also shown
(lines).  It is apparent that the peak value in the restructuring networks is constantly larger than
the plateau value in the corresponding nonrestructuring networks, therefore confirming that the
restructuring of the network happens via cooperative processes that are not present in the
nonrestructuring case.

Overall, we find that the dynamics of our model gel are characterized by restructuring-induced
heterogeneities whose magnitude depends on the length scale probed by the wave vector $q$. A
$q$-dependent dynamical heterogeneity in colloidal gels has indeed been reported in recent light
scattering experiments~\cite{duri2006length,trappe2007_pre}. It is worth mentioning that the
experiments do not show a peak in $\chi_4$ associated to thermally induced fluctuations of the gel
branches (i.e.\ the analogous of our first peak), due to damping of the particle motion through the
solvent. As we will see in the next section, the use of dissipative microscopic dynamics---which are
more appropriate to model a real colloidal suspension---suppresses the first peak in our model as
well, leaving only the peak due to restructuring. The amplitude of the restructuring peak increases
roughly linearly with $q$ (see Fig.~\ref{fig:chi4star}), in accordance with the scaling observed for
the dynamical susceptibility in the experiments reported in Ref.~\onlinecite{duri2006length}.

\section{Comparison of different microscopic particle dynamics}\label{sec:microdyn}
The overall picture we have obtained so far is that the restructuring of the gel networks, although
due to individual bond-breaking events, takes place in a cooperative fashion. We would like at this
point to investigate whether the picture obtained depends on the microscopic dynamics used in the
simulations. In particular we have so far discussed molecular dynamics simulations performed
integrating Newton's equation of motion, whereas the real systems are characterized by damped
microscopic dynamics due to the solvent. Newtonian dynamics allow for a clearer distinction between
the different relaxation regimes (respectively due to the vibrational motion and the overall
relaxation of the network) and has therefore simplified our analysis. In dense glassy systems, it
has been shown that the qualitative behavior of the dynamical susceptibility $\chi_4(q,t)$ does not
change, but there are a few features that depend indeed on the microscopic particle dynamics chosen
for the numerical simulation~\cite{dh2}.

\begin{figure}
  \begin{center}
      \includegraphics[width=1.\columnwidth, clip=true]{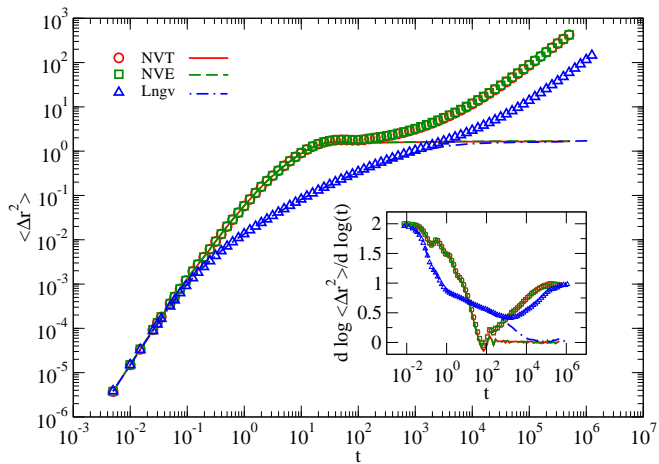}
  \end{center}
  \caption{(Main plot) Dependence of the mean square displacement on the microscopic particle
    dynamics in a restructuring (symbols) and non-restructuring (lines) network with volume fraction
    $\phi=7.5\%$. The temperature is fixed at $T=0.05$. (Inset) Logarithmic derivative of the mean
    square displacement, showing the transition from the initial ballistic regime to the final
    diffusive regime (restructuring network) or the approach to the plateau (nonrestructuring
    network).
    \label{fig:microdyn_msd}}
\end{figure}

We have therefore considered the same gel network and compared three different microscopic particle
dynamics: $NVT$, $NVE$, Langevin. These simulations have been performed for $4000$ particles at
volume fraction $\phi=7.5\%$ and temperature $T=0.05$. The smaller system size is due to the higher
computational cost required by Langevin dynamics; in order to exclude size effects we consider the
same number of particles for all three cases. In the case of $NVT$ dynamics we used the same
thermostat (Nos\'e-Hoover) and equilibration protocol employed for the bigger systems, as detailed
in Section~\ref{sec:numsim}. The $NVE$ simulations were started from a well equilibrated
configuration obtained with $NVT$ dynamics; we checked that the temperature variation was negligible
during the runs at constant energy. In the case of Langevin dynamics we solved for each particle the
equation of motion
\begin{equation}
  m\ddot{\vct{x}} = \vct{F}_c + \vct{F}_f + \vct{F}_r\,,
\end{equation}
where $\vct{F}_c$ is the conservative force derived from the potential~\eqref{equ:poten},
$\vct{F}_f=-m\gamma\dot{\vct{x}}$ is a frictional force, and $\vct{F}_r\propto\sqrt{\kb T m \gamma}$
is a random force due to solvent atoms at a temperature T bumping into the
particle~\cite{kob_binder1998_prl,md}. We chose $\gamma=10.0$ in reduced units, which we found high
enough to make a substantial difference with respect to Newtonian dynamics, yet small enough to
enable the system to reach full relaxation on a time scale accessible to simulations.

In Fig.~\ref{fig:microdyn_msd} we compare the mean square displacement $\langle\Delta
r^2\rangle$~\eqref{equ:msd} for the three microscopic dynamics in a restructuring network (symbols)
and a nonrestructuring one (lines). The curves for $NVT$ and $NVE$ coincide, showing that the two
distinct microscopic dynamics result in virtually identical average particle displacements over
time. In the case of Langevin dynamics, owing to the microscopic dumping of particle motion, after
an initial segment coinciding with the other two curves the dynamics slow down. In the restructuring
network the plateau is significantly smoothened, resulting in a more gradual transition from the
ballistic to the diffusive regime (see also the inset of Fig.~\ref{fig:microdyn_msd}, reporting the
logarithmic derivative $d\log\langle\Delta r^2\rangle / d \log t$). In the nonrestructuring network
the three curves reach at long times the same asymptotic value, showing that the extent of particle
localization, being determined only by the network topology, is independent of the microscopic
dynamics (the temperature being equal).

\begin{figure}
  \begin{center}
      \includegraphics[width=1.\columnwidth, clip=true]{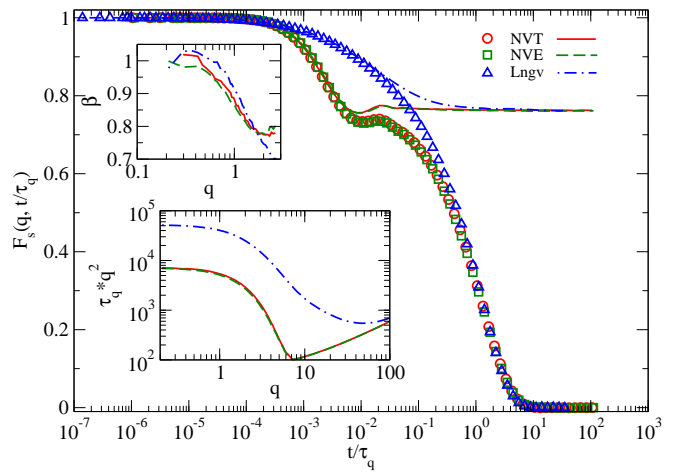}
  \end{center}
  \caption{(Main plot) Incoherent scattering function $F_s(q,t)$ plotted for $q=1.05$ as a function
    of the reduced time $t/\tau_q$ for the different microscopic dynamics in a restructuring network
    (symbols) and in a nonrestructuring network (lines). (Top inset) Stretching exponent of the
    long-time decay of correlations as a function of wave vector. (Bottom inset) Relaxation time,
    multiplied by $q^2$, as a function of wave vector. \label{fig:microdyn_sisf}}
\end{figure}

The incoherent scattering function $F_s(q,t)$~\eqref{equ:sisf} is shown in
Fig.~\ref{fig:microdyn_sisf} for $q=1.05$. Time is rescaled by the structural relaxation time
$\tau_q=\int_0^{\infty} F_s(q,t)\,dt$ to account for the trivial slowing down of the relaxation in
the case of Langevin dynamics. We see again that $NVT$ and $NVE$ dynamics result in equivalent
relaxation patterns. In the restructuring network (symbols) the use of Langevin dynamics causes the
damping of the short-time vibrational motion of the network, leading to the disappearance of the
plateau separating this fast relaxation process from the slower restructuring dynamics. The pattern
of long-time relaxation is instead clearly independent of the microscopic dynamics, since the three
curves fall on top of each other. A similar behavior has been observed in simulations of dense
glassy systems~\cite{kob_binder1998_prl}. As a further confirmation we plot in the upper inset of
Fig.~\ref{fig:microdyn_sisf} the stretching exponent $\beta$ obtained as a function of $q$ from a
fit of a stretched exponential function~\eqref{equ:strexp} to the long-time decay of
$F_s(q,t)$. Although the results are not identical, presumably due to the sensitivity of the result
on the time range chosen for the fit, there is no qualitative difference between the various
microscopic dynamics: the relaxation is exponential at small wave vector, then becomes more and more
stretched ($\beta$ diminishes) by increasing $q$. In the lower inset we plot $q^2\tau_q$ as a
function of $q$: in all cases the dynamics are diffusive ($\tau_q\sim q^{-2}$) at low $q$ and
ballistic ($\tau_q\sim q^{-1}$) at high $q$. In the case of Langevin dynamics the transition is
smoother, and the dynamics ballistic only at very high wave vector because of the microscopic
damping. In the nonrestructuring network (lines in the main plot of Fig.~\ref{fig:microdyn_sisf})
the three curves attain the same asymptotic value, proving again that the extent of particle caging
is not affected by the microscopic dynamics, but instead set by the network topology.

\begin{figure}
  \begin{center}
      \includegraphics[width=1.\columnwidth, clip=true]{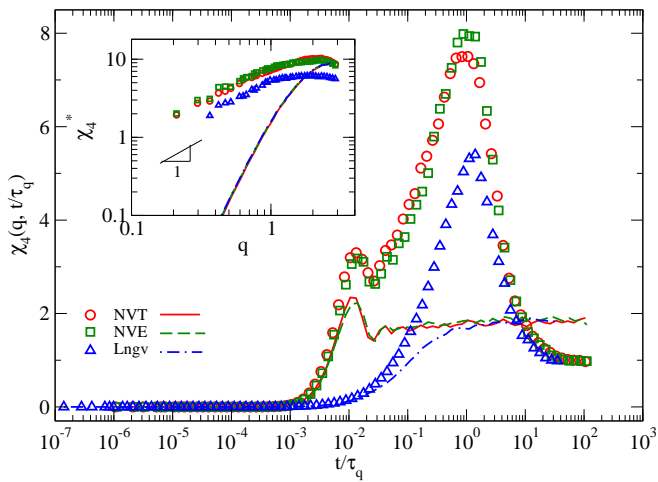}
  \end{center}
  \caption{(Main plot) Dependence of the dynamical susceptibility $\chi_4(q=1.05, t/\tau_q)$ on the
    microscopic dynamics in a restructuring (symbols) and nonrestructuring (lines) network
    ($\phi=7.5\%$, $T=0.05$). (Inset) Amplitude of the restructuring peak of the dynamical
    susceptibility, $\chi_4^*$, as a function of the wave vector $q$ (symbols), compared to the
    plateau value of $\chi_4(q,t)$ in the corresponding nonrestructuring network (lines).
    \label{fig:microdyn_chi4}}
\end{figure}

Finally, we report in Fig.~\ref{fig:microdyn_chi4} the dynamical susceptibility
$\chi_4$~\eqref{equ:chi4} for $q=1.05$ as a function of the rescaled time $t/\tau_q$. Again, $NVT$
and $NVE$ microscopic dynamics produce similar results. In the restructuring network the curves show
two peaks: the first one corresponds to the localization time $t_{\rm loc}$ and is due to network
vibrations; the second one corresponds to the relaxation time $\tau_q$ and arises from network
restructuring (see Section~\ref{sec:coopdyn}). In the case of Langevin dynamics the first peak
disappears completely, presumably due to the damping of the vibrations of the network strands. The
restructuring peak is still present, but its amplitude is smaller than the one of the peaks of $NVT$
and $NVE$ dynamics. We therefore conclude that dynamic fluctuations with \emph{stochastic}
microscopic dynamics are quantitatively different from the fluctuations obtained with
\emph{deterministic} microscopic dynamics, a fact that has already been pointed out for dense glassy
systems in Ref.~\onlinecite{dh2}. In the same work the authors additionally found that at low enough
temperature the fluctuations for $NVT$ and $NVE$ deterministic dynamics also differ (the latter
being roughly equivalent to Langevin dynamics in terms of dynamic fluctuations). In our gel
network---at least at the state point that we have analyzed---the two deterministic dynamics produce
instead similar results; it might be that lower temperatures and/or higher densities are required to
observe a difference as big as the one reported for the dense glass. In the nonrestructuring network
all three curves attain the same plateau value at large time, which is consistent with the results
in Figs.~\ref{fig:microdyn_msd} and \ref{fig:microdyn_sisf}.

In the inset of Fig.~\ref{fig:microdyn_chi4} we plot the amplitude of the restructuring peak of the
dynamical susceptibility, $\chi_4^*$, as a function of the wave vector $q$ (symbols), compared to
the plateau value of $\chi_4(q,t)$ in the nonrestructuring network (lines). Irrespective of the
microscopic particle dynamics, at low wave vector the restructuring process is characterized by a
degree of cooperativity larger than the one that might be explained by simple network vibrations.
We also note that in all three cases the amplitude of the restructuring peak scales roughly linearly
with the wave vector at low $q$, again in accordance with the experimental findings in
Ref.~\onlinecite{duri2006length}.

\section{Summary and conclusions}\label{sec:conclusion}

In this work we have analyzed via numerical simulations the self-assembly, the structure, and the
cooperative dynamics of model colloidal gels. In the spirit of other recent works, our model uses
anisotropic effective interaction to stabilize thin stress-bearing networks at low particle volume
fractions. Starting from a colloidal gas at high temperature, upon lowering the temperature the
particles first aggregate into chains, a process that is well described by a Flory-Huggins type of
mean field theory. At low enough temperature the chains cross-link and percolate, leading to
dynamically arrested networks, i.e.\ gels. Although in reality there is a variety of pathways to
gelation (involving, for instance, arrested spinodal decomposition~\cite{plu_nature}), our assembled
networks capture a few distinctive traits of real colloidal gels. In addition to analyzing the
structure factor, we also observed that the spatial distribution of crosslinks is \emph{not}
homogeneous, but shows spatial correlations ascribable to the interactions stabilizing the
network. This is probably a distinctive feature of physical gels, in which crosslinking is driven by
the same interparticle interactions that lead to the aggregation, whereas in chemical gels
irreversible crosslinking is usually initiated at random locations throughout the sample.  In the
range of temperatures that we have simulated the interparticle bonds are not permanent: thermal
fluctuations favor the breaking of existing bonds and the formation of new bonds. Thus the networks
\emph{restructure} over time, an inherent feature of colloidal gels. Since the persistence length of
the chains in our model is of the order of, or greater than, the average chain length, our networks
have interesting similarities with fiber gels such as semiflexible biopolymer
networks~\cite{fred_prl1,fred_prl2}.

The structural relaxation of the networks as measured by the average particle dynamics follows a
two-step decay pattern reminiscent of glassy dynamics: two relaxation processes taking place on
different time scales are separated by an intermediate plateau. By comparing the dynamics of gels
differing only in the possibility to restructure, we have been able to show that the fast relaxation
process corresponds to fluctuations of the network strands that do not entail any breaking of bonds;
on the contrary, the final decay of correlations is caused by the restructuring of the network. The
plateau separating the two regimes is again reminiscent of the caging effect in dense glasses:
whereas in the glass the caging arises because of the crowding of the particles, in the gel it is a
consequence of the particles being embedded in the network structure through relatively long-lived
bonds. However, escaping from the cage is not to be merely identified with the breaking of the local
bonds connecting a particle with its neighbors; instead, it is a process that involves a much larger
environment, i.e.\ the restructuring of the whole network. As we have shown in
Ref.~\onlinecite{jaderema_prl} this happens in a cooperative fashion because breaking of
interparticle bonds has nonlocal consequences, in the sense that it affects the displacements of
particles located in regions of the gel different from the one where the breaking event originated.

The cooperative nature of the network restructuring translates into a length-scale dependent
relaxation behavior. The final decay of correlations measured by the incoherent scattering function
$F_s(q,t)$ is exponential at low wave vector, but becomes more and more stretched by increasing $q$
in the range of wave vectors that is able to capture the glassy dynamics. In the same range of wave
vectors the dynamical susceptibility $\chi_4(q,t)$ displays in the restructuring gels a pronounced
peak corresponding to the structural relaxation time $\tau_q$, whereas it saturates to a much lower
plateau in the nonrestructuring case.
The amplitude of the restructuring peak scales as $q^{-1}$ at low wave vector.

We have performed simulations with different choices of the statistical ensemble and of the
microscopic particle dynamics: Newtonian dynamics at constant energy ($NVE$) and temperature
($NVT$), and stochastic dynamics at fixed temperature (Langevin). We observed that the
\emph{absolute value} of the restructuring peak of $\chi_4$ when using stochastic dynamics is
different from the ones obtained with Newtonian dynamics, in accordance with earlier observations
made in a dense glass~\cite{dh2}. Nevertheless, the pattern of the long-time relaxation of the gel
is robust with respect to the choice of microscopic dynamics and statistical ensemble: the
stretching exponents, the scaling of the relaxation time with the wave vector, and the
$q$-dependence of the peak amplitude of $\chi_4$ are very similar in the three cases. 
In this respect, although the choice of a damped microscopic dynamics such as Langevin might appear
more relevant for modeling the real system (in which the particles are embedded in a solvent)
Newtonian dynamics has the advantage of being less demanding in terms of computational resources,
and enables as well a sharper separation of the short-time vibrational dynamics from the long-time
restructuring process. Therefore, whenever the interest lies in the long-time properties of the gel
the choice of Newtonian microscopic dynamics is legitimate. 

There is one aspect of the real system that we have completely ignored: hydrodynamics. It has been
shown that many-body hydrodynamic interactions can promote gelation, or lower the colloid volume
fraction threshold for percolation, as compared to their absence; they also appear to influence the
morphology of the particle clusters forming during the aggregation~\cite{tanaka_hydro}.  Although
the results discussed here support the idea that the long-time glassy and cooperative dynamics
should not be quantitatively changed by hydrodynamics, these recent works suggest that hydrodynamics
may significantly affect the kinetic path to gelation, leading to possibly strong differences in the
structure of the gel network, and these changes may in turn eventually affect the microscopic
cooperative processes.  This point requires further investigations.

The dynamics of our model system encompass many of the distinctive traits that have been measured in
real colloidal gels with the use of advanced light scattering
techniques~\cite{duri_pre-2005,duri2006length}: the structural relaxation time extracted by the
\emph{coherent} scattering function shows a ballistic scaling with the wave vector at low $q$
($\tau_q\sim q^{-1}$); in the same range of wave vectors the peak value of the dynamical
susceptibility has a linear $q$-dependence ($\chi_4^*\sim q$); the final decay of correlations after
the plateau is nonexponential, with a $\beta$ exponent decreasing with increasing wave
vector. However, while in our model the decay is \emph{stretched} ($\beta\lesssim 1$), in the
experiments a \emph{compressed} relaxation ($\beta\gtrsim 1$) is observed, a behavior that should be
ascribed to the aging of the material under the effect of internal stresses quenched in the gel
structure during solidification~\cite{bouchaud,cipelletti2003_faraday}. Hence to investigate this
part of the relaxation dynamics, deeply quenched gel configurations should be analyzed. The authors
of Ref.~\onlinecite{duri2006length} propose that the slow relaxation dynamics of a colloidal gel and
the associated dynamical fluctuations result from a series of discrete, instantaneous rearrangement
events taking place inside the gel volume as a consequence of the progressive relaxation of internal
stresses. The average degree of correlation $F(q,t)$ and the related dynamical susceptibility
$\chi(q,t)$ can therefore be quantitatively connected with the number of such events taking place in
a time interval $t$. If in our model gel we identify the elementary rearrangement event with the
breaking of a single bond, then an elementary event can have quite different effects depending on
the local environment of the broken bond~\cite{jaderema_prl}. Hence the structural relaxation of our
gel, although clearly correlated to bond breaking, is not simply determined by the total number of
breaking events taking place in a specified time interval. This suggests that the elementary
rearrangements events to be considered might correspond to processes that are more complex than the
breaking of individual bonds.





\bibliography{gel} 

\end{document}